\documentclass[prl,twocolumn,showpacs,preprintnumbers,amsmath,amssymb]{revtex4}

\usepackage{graphicx}
\usepackage{dcolumn}
\usepackage{bm}

\begin{document}


\title{Evidence for parametric memory and correlation in semiconductor microcavities}

\author{S. Kundermann$^1$, M. Saba$^{1*}$, C. Ciuti$^2$, T. Guillet$^1$, JL. Staehli$^1$, and B. Deveaud$^1$}
\affiliation{$^1$ Institut de Photonique et d'Electronique Quantique, Ecole Polytechnique F\'ed\'erale de Lausanne (EPFL),\\ CH-1015 Lausanne, Switzerland.\\
$^2$  Laboratoire de Physique de la Mati\`{e}re Condens\'{e}e,
Ecole Normale Sup\'{e}rieure, Paris, France.}

\date{\today}

\begin{abstract}
We measured the dynamics of polariton parametric stimulated scattering in semiconductor microcavities, by time-resolving the amplified signal with 250~fs~resolution. Our experiments demonstrate that the stimulation process is considerably delayed with respect to the arrival time of both probe and pump pulses.   This effect is clearly observable in our high quality sample due to the long lifetime of the microcavity polaritons (about 10 ps), and it is in excellent agreement with the model describing the coherent conversion of two pump polaritons into a signal-idler pair. We demonstrate that the non-instantaneous character of the polariton stimulation occurs because the polariton-wave amplification requires the build up of correlated signal-idler polariton fields. 
\end{abstract}

\pacs{78.65.-s, 05.30.Jp, 42.65.-k.}

\maketitle

Matter-wave amplification is a fascinating subject, which is attracting great interest in the physics of  atomic Bose-Einstein condensates \cite{DengFWM,InouyeFWM,Vogels}, but also in the field of semiconductor quantum optics \cite{Yamamotoboser,Lesidang,Rubo,Messin}. In semiconductor microcavities, the particles involved in the coherent collisions are exotic bosons, the so-called microcavity polaritons, resulting from the strong exciton-photon coupling \cite{Weisbuch}. Recently, spectacular amplification phenomena due to stimulated polariton scattering have been experimentally demonstrated \cite{SavvidisPPA,Sabanature,Stevenson,DasbachGain}. Polariton-wave collisions can also be manipulated by judiciously phase-locked excitation pulses, which control the emission over the whole momentum space \cite{KundermannCC}. 

Particularly interesting is the parametric nature of the scattering. The basic underlying process is the coherent conversion of two pump polaritons into a signal-idler polariton pair conserving total energy and in-plane momentum thanks to the very peculiar shape of the dispersion relation of polaritons \cite{CiutiOPA}. Following this description, the signal idler polariton pairs should be entangled \cite{Schwendimannstatistics,SavastaPRL}. Squeezing of the polaritons has been recently been observed in a geometry where signal and idler are degenerate which confirms the possibility to achieve quantum correlation by this non-linear effect \cite{Baas}. The process is initially started by providing a small incident signal occupation with a weak resonant probe laser pulse. The dynamical aspects of the parametric scattering process are therefore particularly interesting, and very seldomly studied \cite{Erland2}. In contrast to the stimulation processes which occur for example in a laser, where the scattering rate into the one-particle final state is directly proportional to its occupation, the situation in the polariton system is more complicated. The stimulation of the parametric scattering involves the entire final states of the conversion process, i. e. the signal-idler pairs \cite{Schwendimannstatistics}. The behavior should differ largely from the dynamics observed in the case of a laser (stimulation from an incoherent reservoir) and be similar to what occurs in a conventional optical parametric oscillator (coherent wave mixing). However, in our microcavity the parametric process involves real mixed states, whose excitonic part also influences the dynamics.

The investigation of the polariton stimulated dynamics is a challenging task, because the polariton signal temporal dynamics needs to be measured with a time-resolution much shorter than the relaxation times of the polariton quasi-particles. 
In order to achieve this goal, we have implemented an up-conversion set-up with 
sub-picosecond resolution ($\approx 0.25$ ps) to directly measure the temporal dynamics of polariton signal amplification. The experiments have been performed on a very high quality microcavity sample with very slow decoherence, whose polariton radiative lifetime exceeds $10$ ps.  
To our knowledge, our experiments demonstrate very directly for the first time the non-instantaneous nature of the polariton parametric stimulated scattering. Our claim is further susbstantiated by an excellent agreement with the predictions given by the polariton parametric amplifier model \cite{CiutiOPA} and correlation effects between signal and idler are evidenced.

\begin{figure}[t!] \centerline{\includegraphics[scale=1]{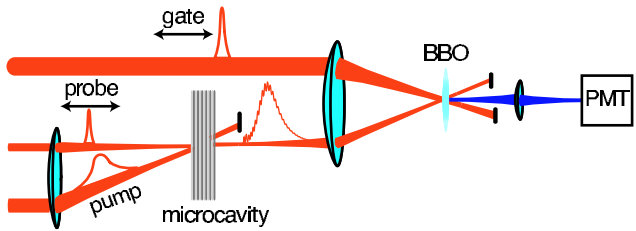}}
\caption{Pump-probe upconversion set-up.}
\label{Upconversionsetup}
\end{figure}

The sample is a single quantum well $\lambda$-cavity held in a helium-bath cryostat at a
temperature of 2 K. The quality of the sample can be assessed by either the linewidth of the normal modes, or the decay of the lower polariton polarization after pulsed excitation. The linewith is of the order of $100{\mu}eV$ and the photon lifetime in excess of 10 ps.
The set-up is an angle-resolved pump and probe apparatus combined with an 
up-conversion detection (see Fig.\ref{Upconversionsetup}). 
The probe beam impinges onto the sample at normal
incidence, whereas the pump arrives at an off-normal angle of about 10 deg.,
which is the so-called magic angle for producing parametric scattering for this microcavity. 
Both pump and probe are pulsed with 80 MHz repetition rate; 
the pump is spectrally filtered to selectively excite the lower polariton branch ($\simeq$1
meV spectral width and $\simeq$1 ps temporal duration), while the weak
probe keeps a broader spectrum ($\simeq$ 15 meV) and shorter duration
($\simeq$150 fs). 
For the upconversion detection, the transmitted probe signal is focused onto a BBO crystal
together with a 150 fs non collinear gating pulse. The cross-correlation beam generated in
the crystal is then selected by a diaphragm and detected with a bi-alkali photomultiplier, so that the
shape of the probe transmission in real time is reconstructed with a resolution
around 250~fs by scanning the delay of the gate pulse. All experiments are carried out far from the regime of saturation, and clearly in the strong coupling regime.

\begin{figure}[!t] \centerline{\includegraphics[scale=0.92]{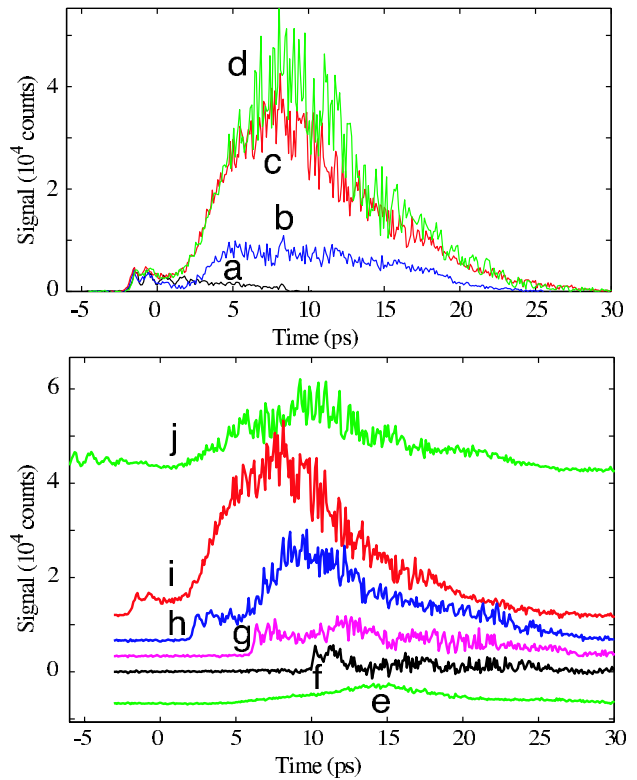}}
\caption{Temporal evolution of the signal at $k=0$ as measured in the experiment. The origin of the time scale corresponds to the arrival of the pump. In the upper panel the pump intensity was varied. The pump intensity is 0 (a), $30W/cm^{2}$ (b), $60W/cm^{2}$ (c), and $80W/cm^{2}$ (d). On the lower panel measurements employing different pump to probe delays are shown. The pump intensity is  $60W/cm^{2}$. The signal of the pump luminescence 2 (e) has been subtracted. The probe hits 5.5ps (j) and 1.5ps (i) before, and 2.5 (h), 6.5 (g), and 10.5ps (f) after the pump. The probe intensity for all measurements is $0.3W/cm^{2}$. }
\label{experimentalfigure}
\end{figure}

The upper panel of Fig.~\ref{experimentalfigure} shows the measured temporal evolution of the stimulated parametric scattering for different pump densities. The origin of the time scale corresponds to the arrival time of the pump \cite{lasernoise}. The probe hits the sample 1.5\,ps before the pump (similar results are obtained for other delays between the pump and the probe). 
Curve (a) shows the temporal evolution of the transmitted probe signal in absence of the pump. Due to the simultaneous excitation of the upper and lower polariton branch the temporal evolution exhibits oscillations due to quantum beats between lower and upper polariton. This signal decays with the expected cavity lifetime.
 When the pump density is adjusted above threshold (trace 2 (b)), the probe signal first starts decaying before a rise due to the stimulated scattering amplification takes place. This rise is significantly delayed with respect to the arrival of the pump pulse. When the pump intensity is further risen, the scattering even saturates (2c, 2d).  As previously noticed the process is significantly delayed with respect to the arrival of the pump. 

Curves (b)-(d) in Fig.~\ref{experimentalfigure} show the delayed build up of the scattered signal. To get a better insight into this behavior, the experiments were performed for different pump to probe delays as shown on the lower panel. The pump intensity is $60W/cm^{2}$, as for trace (c). 
For an easier reading of the data, the weak signal arising from  the "spontaneous" parametric luminescence of the pump measured with the pump alone (trace 2 (e)) has been subtracted in each case. 
In all cases, the stimulated  emission is strongly delayed with respect to the incident probe pulse. When the probe hits the sample first ($-5.5ps$ (j) and $-1.5ps$ (i)) the transmitted signal first decays as expected and then gets amplified, but only with some delay with respect to the arrival of the pump. Curve (i) peaks at an earlier time than (j) because the polariton intensity at $t=0$ and $k=0$ is higher. 
Even when the probe hits the sample after the pump, the stimulated emission is clearly delayed (traces (f), (g), and (h)).
These non trivial dynamics are somewhat astonishing as
they cannot be explained using a simple exponential dependence
of the stimulation. This demonstrates that the often assumed stimulated process following (N+1) where N is the sole population of the k=0 polariton state is far too simple. To understand in detail the dynamical behavior we now turn to some theoretical considerations.

The basic process for this kind of polariton amplification is considered to be the parametric conversion of two pump polaritons with wave-vector $k_p$ into a signal-idler pair, with conservation of both total energy and momentum. In its simplest version, the scattering in the system can be described by three modes only, at $k=0$, $k=k_p$ and $k=2k_p$ in a factorized mean-field approximation \cite{meanfield,CiutiOPA}. 
Since the parametric scattering involves states on the lower polariton dispersion, it is sufficient to look at the lower polariton branch only to understand the basic underlying dynamics. 
The use of proper polariton states for the lower branch allows then to write the dynamics of the system as a set of three coupled differential equations. 

\begin{eqnarray}
i \hbar
\frac{\partial P_{ 0}}{\partial t}=
(\tilde{E}_{ 0} - i\gamma_0)P_{0}
 + E_{int} P_{2 {k}_p}^{\star}
P_{{k}_p}^2
+ F_{0}(t) ~\label{probe} \\
i\hbar
\frac{\partial P_{k_p}}{\partial t}=
(\tilde{E}_{k_p}-i\gamma_{k_p})P_{k_p}
+2E_{int}P^{\star}_{k_p}
P_{0}P_{2{k}_p}+
F_{k_p}(t) ~\label{pump} \\
i\hbar \frac{\partial P_{2 {k}_p}}{\partial t}=
(\tilde{E}_{2k_p}-i\gamma_{2k_p}) P_{2k_p}
+ E_{int}~P_{0}^{\star}
P_{k_p}^2 ~ \label{idler}
\end{eqnarray}

where $P_0$, $P_{k_p}$, $P_{2k_p}$ are the mean lower
polariton fields for signal, pump, and idler wave-vectors respectively and 
$E_{int}$ is the coupling energy due to polariton-polariton scattering potential.
The quantities $P_{0,k_p,2k_p}$ are the rescaled polarizations 
$P_k(t)=\frac{\lambda _X}{\sqrt A}\langle p_ k\rangle (t)$
($p_ k$ are the polarizations). 
On the rhs of each of the three equations, the first term is the 'free term',
{\it i.e.} the evolution of the polarization of non-interacting polaritons
having energy $\tilde{E}(k)$. The homogeneous k-dependent linewidth $\gamma_{k}$ determines an exponential decay rate of the signal ($\gamma_{k}$ is determined by the losses of polaritons through
the cavity mirrors and also includes non-radiative losses). Note that the
energy $\tilde{E}(k)$ includes a renormalization due to the polariton-polariton
interaction and therefore is slightly higher than the energy $E(k)$ calculated
in the absence of pump polaritons. The second term is the parametric 
scattering rate: for the signal (idler) it is proportional to the square of the pump
polarization ({\it i.e.}, proportional to the pump intensity) and to the
conjugate of the idler (signal) polarization. The symmetry between signal and
idler equations reflects the fact that for each polariton scattering from the
pump down to the signal there is another one scattering up to the idler and vice versa. 
This is the origin of the correlation between signal and idler polaritons.
The
last term in the first two equations ($F_{0,k_p}(t)$) corresponds to the external driving 
electric field, {\it i.e.}, the amplitude of pump and probe laser pulses.

 A crucial physical consequence of parametric scattering is that the rate of scattering from the pump into the signal mode is proportional to the idler field. More directly, we can write

\begin{equation}
\left (\frac{\partial{P_0}}{\partial   t} \right) _{par}= \frac{E_{int}} {i  \hbar}
P^{\star}_{2k_p}(t)  P^2_{k_p}(t),
\label{rate}
\end{equation}

The dynamical equation for the idler polariton field can be formally solved, giving the following 
analytical result :

\begin{equation}
 P_{2k_p}(t) = \frac{E_{int}} {i  \hbar}
\int_{-\infty}^{t} dt'
P^{\star}_{0}(t)  P^2_{k_p}(t') 
e^{-\frac{i}{\hbar} (E^{LP}_{2k_p}-i \gamma_ {2k_p})(t-t')},
\label{idlerrate}
\end{equation}

where $E^{LP}_{2k_p} $ is the idler polariton energy and $ \gamma_ {2k_p}$ the corresponding
broadening.
By inserting Eq. (\ref{idler}) into Eq. (\ref{rate}), we can get the expression for the signal parametric scattering rate in terms of the signal and pump polariton fields only, namely :

\begin{equation}
\left (\frac{\partial{P_0}}{\partial   t} \right) _{par}= 
\left (\frac{E_{int}} {i  \hbar} \right)^2 \int_{-\infty}^{t} dt' K_{par}(t,t') P_0(t'),
\label{memory}
\end{equation}
where the {\it parametric memory} kernel reads
\begin{equation}
K_{par}(t,t') = 
P^{\star 2}_{k_p}(t')  
e^{+\frac{i}{\hbar} (E^{LP}_{2k_p}+i \gamma_ {2k_p})(t-t')}
P^2_{k_p}(t)  .
\label{kernel}
\end{equation}

Eq. (\ref{memory}) shows that the rate of pump scattering into the signal mode is proportional to the signal field itself, i.e.  the scattering is stimulated. However, the stimulation is not instantaneous, because the stimulated scattering rate depends on the value of the signal field at all times $t' <t$.
This memory effect takes place, because the parametric stimulation requires the 
coherent and correlated build-up of the signal and idler fields. 
Instantaneous stimulation is recovered only when the parametric memory kernel  \cite{kernelref}
$K_{par}(t,t')  \propto \delta(t-t')$. From Eq. (\ref{kernel}), we understand that in order to experimentally detect the parametric memory effect, the time-resolution of the 
measurement must be much shorter than the decoherence times of the polariton fields.

\begin{figure}[!t] \centerline{\includegraphics[scale=0.49]{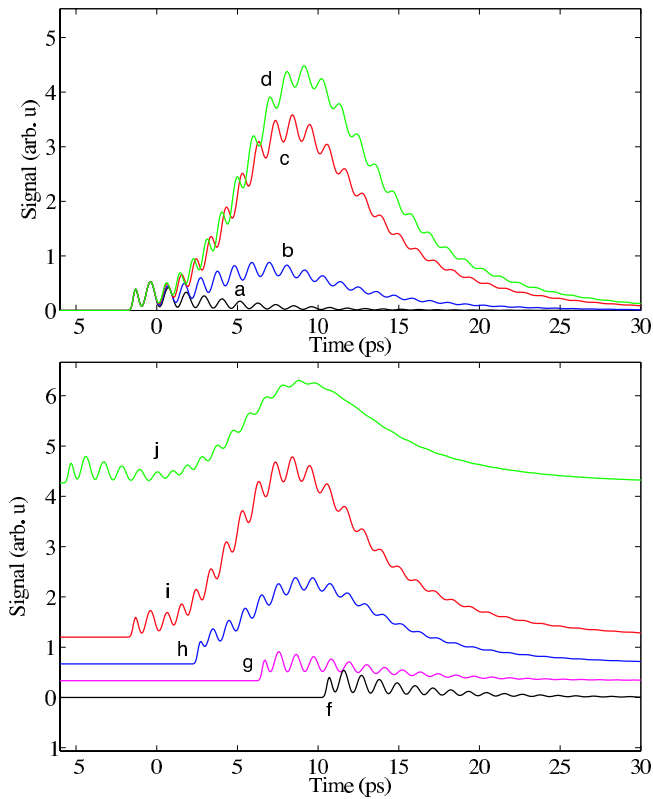}}
\caption{Simulation of the temporal evolution of the signal emission at $k=0$. The pump and probe density has been fitted by a global conversion factor between incident pump power and the respective density, whereas the pump probe delays are the same as in Fig.~\ref{experimentalfigure}. The employed exciton and cavity linewidth is 0.1meV.}
\label{simulationfigure}
\end{figure}

In order to compare the measured dynamics, we have solved numerically the nonlinear equations for the polariton parametric amplifier. Since the short probe pulses also excite the upper polariton branch, we have extended equations 1-3. The calculations have been performed in the exciton-cavity photon basis, so that the calculation accounts for the lower and the upper polariton branch. 
As previously shown for the analytic derivation of the memory kernel (in the lower polariton basis), the same behavior can be found again in the numerical equations. First the external probe laser field creates a polariton occupation in the signal state. This polariton occupation together with the pump polaritons result in a build up of the idler polarization. Since the signal and idler equations are symmetric, the established idler polarization stimulates the pump polaritons to scatter to the signal state and the signal build up is delayed and non mono-exponential. 

Fig.~\ref{simulationfigure} shows the simulation of the dynamics using the model of the parametric amplifier as discussed before. The pump and probe densities 
have been adjusted according to the external pump and probe power by using a global conversion factor and the exciton and cavity linewidth have been inserted according to previous works on similar samples \cite{Stanleyquench}. The pump to probe delays are the same as in the experiment (Fig.~\ref{experimentalfigure}). The agreement with the experimental data is very good. All the features observed in the experiment (Fig.~\ref{experimentalfigure}) are reproduced, even the saturation at the highest pump level. A very important observation is made looking at curves 2 (h) and 3 (f-h): the oscillations due to the LP-UP beating remain even after the pump has hit at t=0, which demonstrates that we are in the strong coupling regime in all cases. However the beat oscillations persist much longer in the simulation than in the experiment because of the absence of additional dephasing mechanisms for the upper polariton in the model \cite{dephasingmechanism}. 
The saturation of the signal intensity observed for curves \ref{experimentalfigure} (b, d) and \ref{simulationfigure} (b, d) is due to the depletion of the pump polariton reservoir because the scattering rate out of the pump reservoir increases with the square of the pump intensity (see equation \ref{kernel}) \cite{KundermannOECS03}.

\begin{figure}[!ht] \centerline{\includegraphics[scale=0.49]{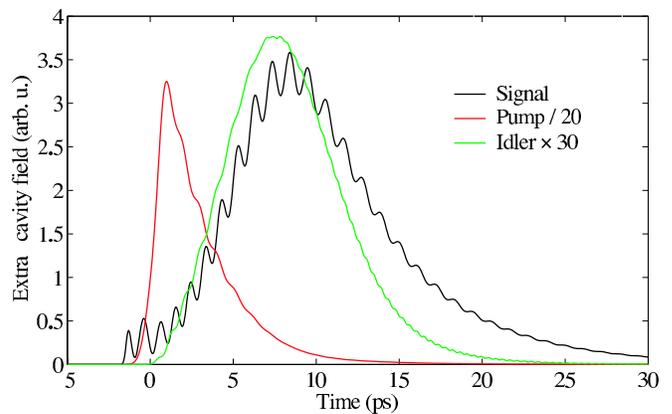}}
\caption{Simulation the extra cavity field of signal, pump, and idler versus time.}
\label{SPIfigure}
\end{figure}

The simulation also gives access to the pump and idler temporal evolution. To further inspect the dynamical interplay of signal, pump, and idler we look at the evolution of the three modes. In this simulation, which is depicted in Fig.~\ref{SPIfigure}, the probe hits the cavity again 1.5\,ps before the pump. 

The idler starts to build up when the pump hits and the signal follows the evolution of the idler after its initial decay. 
The simultaneous build up of signal and idler demonstrates their mutual correlation. 
This shows that even in the mean-field approximation a signal idler correlation appears in spite of the fact that all the field operators have been factorized. In Fig. 2~(e) we observe that the signal due to the "spontaneous" parametric luminescence of the pump needs much more time to build up. The "seed" for this signal is not the initial probe polariton occupation but the vacuum fluctuations of signal and idler modes (a model accounting for that can be found in \cite{Schwendimannstatistics}). Since the effect of the fluctuations starts to build up the signal much later (see curve 2 (e)), their effect can be neglected for the pump probe experiment, in which the correlations are built up very quickly due to the presence of the probe polaritons. 
The quantum beats between upper and lower polariton resonance appear very clearly in the signal, they also give rise to faint oscilltations in the pump and idler branches. 
 
In conclusion, we have been able to observe the non-instantaneous nature of the polariton parametric scattering in semiconductor microcavities by time resolved pump-probe experiments.  The excellent quality of the sample together with the high time resolution of the experimental setup allow to report this effect for the first time. The measurements are in excellent quantitative agreement with a mean field theory and it is shown that the stimulation takes place due to the mutual correlation of signal and idler. The demonstrated signal idler correlation makes the polaritons a promising system for emission of entangled photons and for quantum information processing. 

We thank U. Oesterle, R. Stanley, R. Houdr\'e, and M. Illegems who provided us with the excellent microcavity sample. Furthermore we would like to acknowledge fruitful discussions with A. Baas, R. Idrissi,  A. Quattropani, V. Savona, and P. Schwendimann. The present work has been supported by the Swiss National Science Foundation. 

$^\star$ Present address: Center for Ultracold Atoms, MIT, Cambridge, Ma, USA.

\bibliographystyle{prsty}

\begin{thebibliography}{10}

\bibitem{DengFWM}
L. Deng {\it et~al.}, Nature {\bf 402},  641  (1999).

\bibitem{InouyeFWM}
S. Inouye {\it et~al.}, Nature {\bf 414},  731  (2001).

\bibitem{Vogels}
  J. M. Vogels {\it et~al.}, Phys. Rev. Lett. {\bf90},  030403  (2003).

\bibitem{Yamamotoboser}
Y. Yamamoto, Nature {\bf 405},  629  (2000).

\bibitem{Lesidang}
Le Si Dang {\it et~al.}, Phys. Rev. Lett. {\bf 81},  3920  (1998).

\bibitem{Rubo}
Y. G. Rubo {\it et~al.}, Phys. Rev. Lett. {\bf 91},  156403  (2003).

\bibitem{Messin}
G. Messin {\it et~al.}, Phys. Rev. Lett. {\bf 87},  127403  (2001).

\bibitem{Weisbuch}
C. Weisbuch {\it et~al.}, Phys. Rev. Lett. {\bf
  69},  3314  (1992).

\bibitem{SavvidisPPA}
P.~G. Savvidis {\it et~al.}, Phys. Rev. Lett. {\bf 84},  1547  (2000).

\bibitem{Sabanature}
M. Saba {\it et~al.}, Nature {\bf 414},  731  (2001).

\bibitem{Stevenson}
R.~M. Stevenson {\it et~al.}, Phys. Rev. Lett. {\bf 85},  3680  (2000).

\bibitem{DasbachGain} 
G. Dasbach {\it et~al.}, Phys. Rev. B {\bf62},  13076  (2000).

\bibitem{KundermannCC}
  S. Kundermann {\it et~al.}, Phys. Rev. Lett. {\bf91},  107402  (2003).

\bibitem{CiutiOPA}
C. Ciuti {\it et~al.}, Phys. Rev. B {\bf
  62},  R4825  (2000).

\bibitem{Schwendimannstatistics} 
P. Schwendimann {\it et~al.}, Phys. Rev. B {\bf 68},
  165324 (2003).

\bibitem{SavastaPRL}
  S. Savasta	 {\it et~al.}, Phys. Rev. Lett. {\bf90},  096403  (2003).

 \bibitem{Baas}
 J.Ph. Karr {\it et~al.}, to appear. 

\bibitem{Erland2}
   J. Erland {\it et~al.}, Phys. Rev. Lett. {\bf 86},  5791  (2001).

\bibitem{lasernoise}
The strong noise fluctuations in the measurements are due to fluctuations of the incident laser power. These fluctuations are strongly amplified by the nonlinear characteristics of the scattering. 

\bibitem{meanfield} 
Mean-field approximation in this case means that in the scattering terms in eq. 1-3 the product of the mean values is calculated and not the mean value of the product of the operators. 

\bibitem{Stanleyquench}
  R. Stanley {\it et~al.}, Solid State Comm. {\bf 106},  485  (1998).

\bibitem{kernelref}
Th. \"Ostreich {\it et~al.}, Phys. Rev. Lett. {\bf83},  3510 (1999).

\bibitem{dephasingmechanism} 
E. g., the scattering of an upper polariton with an acoustic phonon into a high momentum exciton state.

\bibitem{KundermannOECS03}
S. Kundermann {\it et~al.} to appear in Phys. Stat. Sol., special issue OECS'03.

\end{thebibliography}

\end{document}